\documentclass[]{aa}
\usepackage{array}
\usepackage{graphicx}
\usepackage{epsfig}

\usepackage{txfonts}

\newcommand{\bv}{(B-V)}
\newcommand{\ub}{(U-B)}
\newcommand{\vi}{(V-I)}

\begin{document}
\title{The VIRMOS deep imaging survey: III. ESO/WFI deep U-band imaging
of the 0226-04 deep field. 
\thanks{Based on observations carried out at the ESO MPI 2.2m telescope located
at La Silla, Chile.}
}

\author{M. Radovich\inst{1}, M. Arnaboldi\inst{1,2}, V. Ripepi\inst{1},
M. Massarotti\inst{1}, 
H. J. McCracken\inst{3,7}, Y. Mellier\inst{4,5},
E. Bertin\inst{4,5}, \\G. Zamorani\inst{7,8}, \\ 
C. Adami\inst{6}, S. Bardelli\inst{7},  O. Le F\`evre\inst{6}, 
S. Foucaud\inst{6}, B. Garilli\inst{9}, R. Scaramella\inst{10}, 
G. Vettolani\inst{7}, A. Zanichelli\inst{8}, \\E. Zucca\inst{7}}

\institute{I.N.A.F.,Osservatorio Astronomico di Capodimonte, Via Moiariello
16, 80131 Napoli, Italy 
\and I.N.A.F.,Osservatorio Astronomico di Pino
Torinese, Via Osservatorio 20, 10025 Torino, Italy 
\and Universit\`a degli Studi di Bologna, Dipartimento di Astronomia, 
Via Ranzani 1, 40127 Bologna, Italy
\and Institut d'Astrophysique de Paris, 98bis Bd. Arago, 75014 Paris, France 
\and Observatoire de Paris, LERMA, 61 Avenue de l'Observatoire, 75014 Paris,
France 
\and  Laboratoire d'Astrophysique de Marseille, Traverse du Siphon, 
13376 Marseille Cedex 12, France
\and I.N.A.F., Osservatorio Astronomico di Bologna, Via Ranzani 1, 
40127 Bologna, Italy 
\and Istituto di Radioastronomia del CNR, Via Gobetti 101, 40129 Bologna, 
Italy
\and Istituto di Astrofisica Spaziale e Fisica Cosmica del CNR, 
Via Bassini 15, 20133 Milano, Italy 
\and I.N.A.F., Osservatorio Astronomico di Roma, Via Osservatorio 2, 
00040 Monteporzio Catone (Roma), Italy
}

\offprints{radovich@na.astro.it}

\date{Received XXX; accepted XXX}

\abstract{
In this paper we describe the $U-$band imaging of the F02 deep field,
one of the fields in the VIRMOS Deep Imaging Survey. The
observations were done at the ESO/MPG 2.2m telescope at La Silla (Chile) using
the 8k $\times$ 8k Wide-Field Imager (WFI).  
The field is centered  at $\alpha$(J2000)=02$^{h}$26$^{m}$00$^{s}$ and
$\delta$(J2000)=-04$^{\circ}$30\arcmin00\arcsec, 
the total covered area is 0.9 $\deg^2$ and the limiting magnitude (50\%
completeness) is  $U_{AB} \sim 25.4$ mag.
Reduction steps, including astrometry, photometry and catalogue
extraction, are first discussed. The achieved astrometric accuracy (RMS) is $\sim
0.2\arcsec$ with reference to the $I-$band catalog and $\sim 0.07\arcsec$ 
internally (estimated from overlapping sources in different exposures). 
The photometric accuracy
including uncertainties from photometric calibration, is $< 0.1$ mag.
Various tests are then performed
as a quality assessment of the data. They include: (i) the color distribution
of stars and galaxies in the field, done together with the $BVRI$ data
available from the VIMOS survey; (ii) the comparison with previous
published results of $U-$band magnitude-number counts of galaxies.
}

\authorrunning{M. Radovich et al.}

\titlerunning{The VIRMOS U deep field}

\maketitle

\begin{figure}
\begin{center}
\includegraphics[%
  width=6cm,
  angle=270]
  {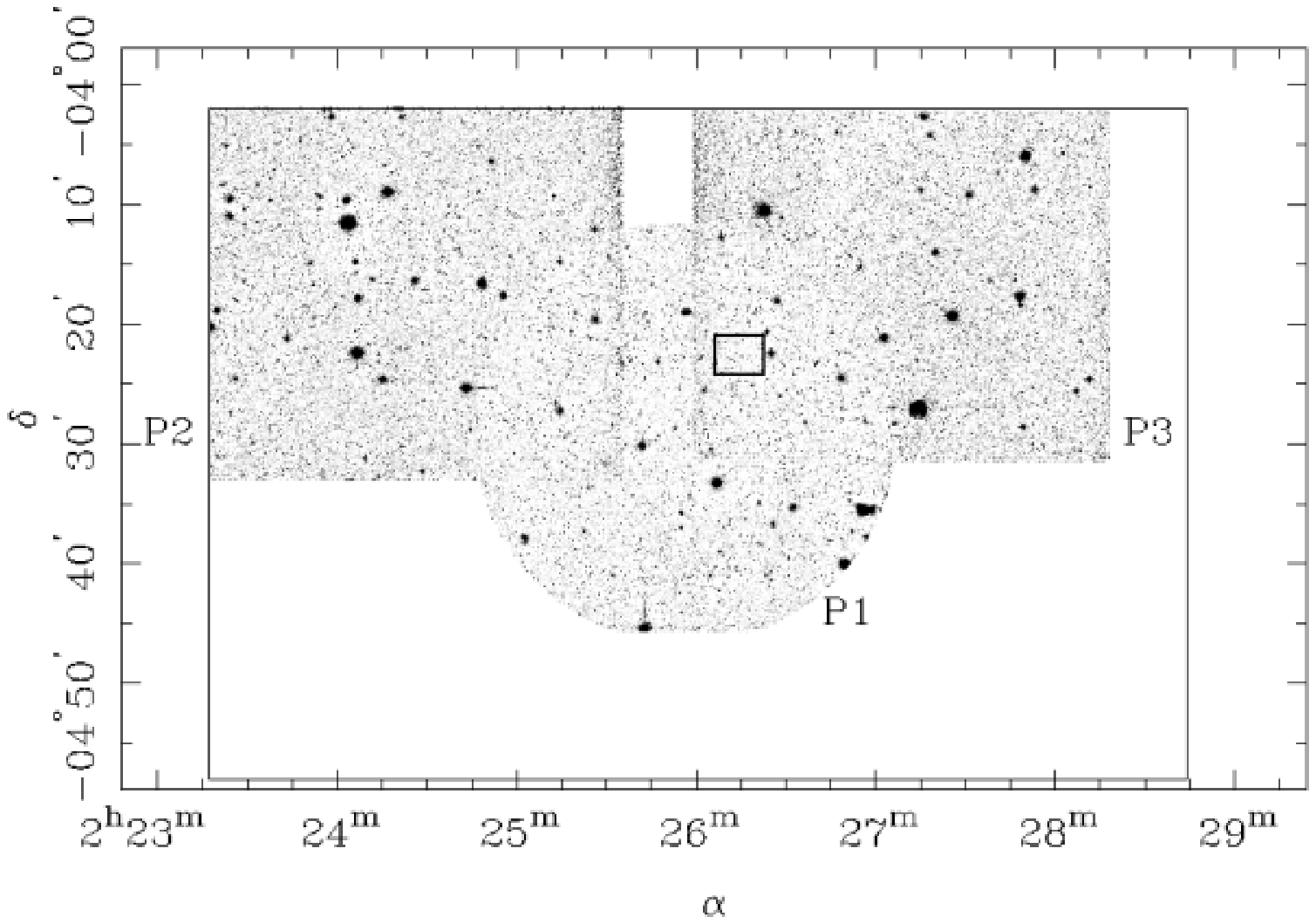}
  \includegraphics[%
  width=6cm,
  angle=270]
  {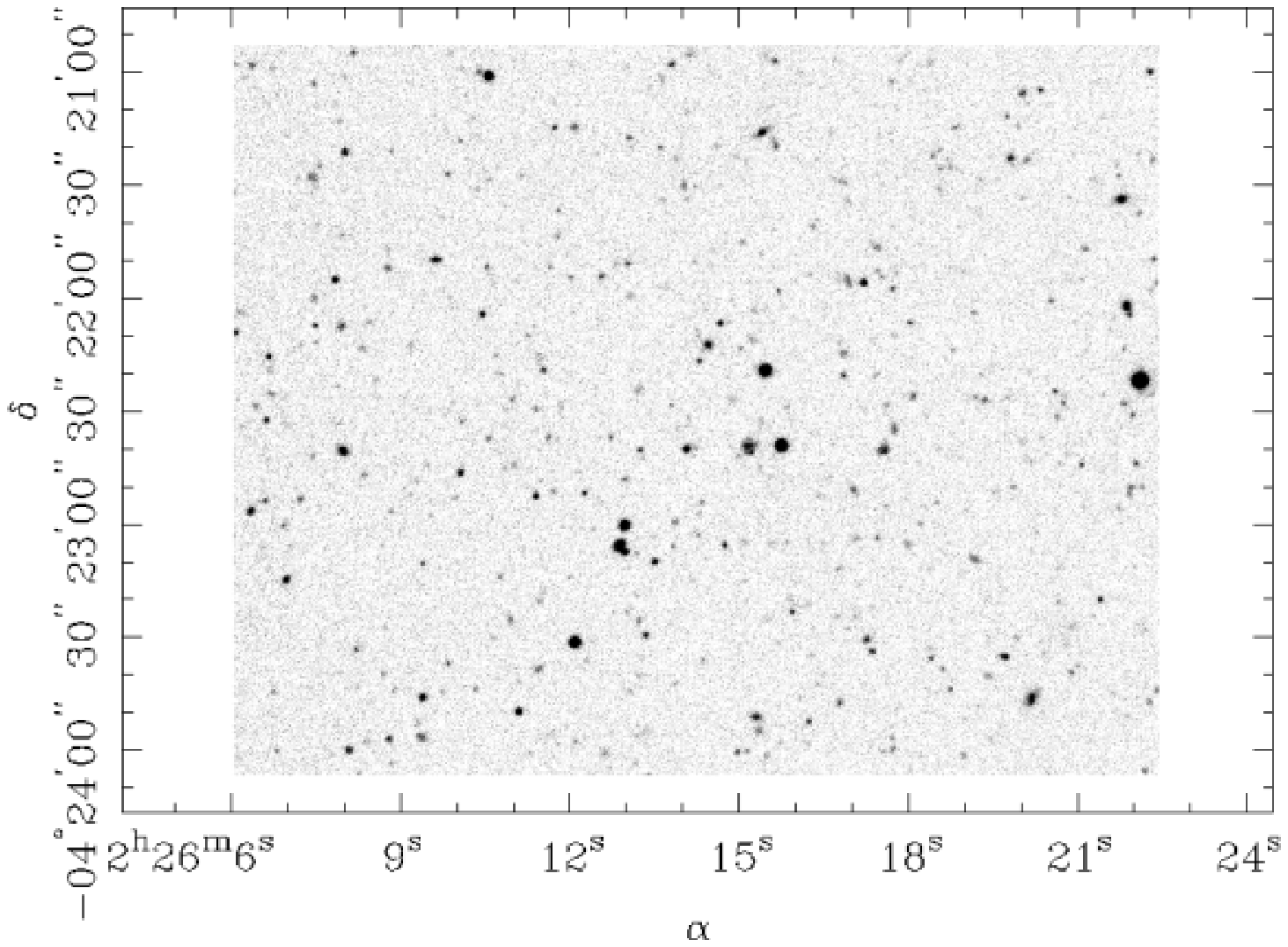}
 \end{center}
\caption{The VIRMOS F02 deep field. The images of the three pointings (P1, P2
and P3) are here combined for display purposes only: they have been treated 
separately in data reduction and catalog extraction.
The full field is displayed in the {\em upper} plot. The outer solid--line
rectangle shows the position of the
$\sim 1.2\deg^2$  field taken in the $BVRI$ bands. Overlaid 
are the images of the three $U-$band pointings; masked areas at the borders 
of the images were removed. Note the circular shape of the
P1 pointing, due to vignetting in the Loiano filter. 
The region in the inner box, where P1 and P3 overlap, 
is the one zoomed in the {\em bottom} plot: the accuracy of the astrometry can be seen 
from the absence of double sources.}

\label{Fig:field}
\end{figure}

\section{Introduction}

The VIMOS imaging survey (Paper I) represents the preparatory step of the deep
redshift survey, which is now being carried out with the VIMOS spectrograph
at the VLT UT3 at Paranal, Chile, by the VIMOS consortium. This preparatory
multi-wavelength imaging survey has been done  at the Canada-France 
Hawaii Telescope (CFHT) for the $BVRI$ bands (Le F\`evre et al. 2003, 
Paper I hereafter; McCracken
et al. \cite{paperII}, Paper II hereafter), and at the ESO NTT telescope 
for the K imaging (Iovino et al. 2003, in prep). The near ultraviolet part
of this survey was carried out at the ESO MPI 2.2m telescope
with the WFI 8kx8k camera, in the framework of an approved ESO Large
program (164.0-0089) scheduled since Period 63 in 1999, both in visitor 
and service mode.

The survey  aims at covering $\sim 16 \deg^2$
in $UBVRI$ with a 5 $\sigma$ limiting magnitude $I_{AB}\sim24.5$, and a 
smaller $1.2 \deg^2$ area reaching $I_{AB} \sim 25$, i.e. the 
VIMOS deep field. 
In the $U-$band, expected 5 $\sigma$ limiting magnitudes are 
$U_{AB} \sim 24$ mag for the wide survey, $U_{AB} \sim 26$ mag for 
the deep survey. In total the survey contains
over $10^{6}$ galaxies in five colors and represents a major
advance over other previous deep multicolor surveys, probing  
structures on scales of $\sim20h^{-1}$ Mpc at $z\sim1$.

Surveys in the radio continuum (Bondi et al. \cite{bondi})
and with XMM (Pierre et al. \cite{pierre}) were carried out on the
deep field of the VIMOS imaging survey. 
The catalogs from the $U$ and $BVRI$ deep field
are also foreseen to be used for the optical identification of the
radio continuum survey (Ciliegi et al. in prep).

Until the availability of data  from ground-based 
(e.g. the CFHLS and VST surveys) or space (e.g. the {\sc GALEX} project) surveys, 
wide-field $U-$band data with medium depth and large sky coverage are not available
yet. 
The $U-$band is essential to study the star formation properties of
field galaxies in the nearby universe, and is also
particularly suited to identify starburst galaxies, AGN, and QSO at
moderate redshift. 
These aspects, as well the determination of the $U-$band
luminosity function and the analysis of the morphological properties of 
galaxies compared to other bands, will be the subject of separate papers.

Compared to other existing $UBVRI$ surveys, the VIRMOS F02 deep field 
offers a good compromise between covered area and  depth. The ESO imaging survey (EIS)
in the Chandra Deep Field South (Arnouts et al. \cite{arnouts}) is comparable in depth to our 
F02 deep field but covers a smaller area ($\sim0.25\deg^2$). 
The Canada-France Deep Field survey (McCracken et al. \cite{cfdf}) consists of four independent fields of 
$\sim 0.25\deg^2$ each, and the $U-$band limiting  magnitude is $\sim$ 1 mag 
deeper than the VIRMOS deep field. 
The Combo-17 survey (Wolf et al. \cite{wolf}) covers three fields with a total area 
$\sim 0.78\deg^2$ and  is comparable in depth to the F02 deep field in the $U-$band. 
Deeper surveys (e.g. the  William Herschel Deep Field, Metcalfe et al. \cite{metcalfe} 
or more recently the FORS Deep Field, Heidt et al. \cite{heidt}) were  
carried out on smaller 
areas ($\sim 7 \times 7 \rm{arcmin}^2$) only.  

In this paper we will describe the observations, photometric calibration,
catalogue extraction and validation for the deep field. The calibration
steps are strongly related to those followed in the case of the overlapping
deep $BVRI$ filters, which are described in Paper II. We refer to this
paper for a detailed discussion of calibration issues when they are
the same and emphasize those aspects that are peculiar to the $U-$band
data.

The paper is organized as follows. In Section 2 we describe the observations
for the deep pointings, the observing strategy and data reduction issues
(pre-reduction, astrometric and photometric calibration). Catalog extraction
is discussed in Section 3. Validation of the catalogs through a number
of tests (comparison of stellar and galaxy colors with simulations,
number counts) is done in Section 4. Conclusions
are drawn in Section 5.

\section{Observations and data reduction}

\subsection{Observations}

The $U-$band deep pointing was observed during several observing runs since
1999 with the aim of covering $\sim 1 \deg^{2}$ of the multiwavelength
deep field planned within the VIMOS deep redshift survey. This deep
field is centred at $\alpha$(J2000)=02$^{h}$26$^{m}$00$^{s}$ and
$\delta$(J2000)=-04$^{\circ}$30\arcmin00\arcsec (see Fig. \ref{Fig:field})
and observations were carried out with the Wide-Field Imaging (WFI)
mosaic camera mounted on the ESO MPI 2.2m telescope at La Silla,
Chile. The camera was mounted at the Cassegrain focus of the telescope,
giving a field of view of 34$\times$33 arcmin. It consists of a mosaic
of 8 CCD detectors with narrow inter-chip gaps, yielding a filling
factor of 95.9 \% and a pixel size of $0.24\arcsec$. The WFI CCDs have
a read-out noise of 4.5 e$^-$ pix$^{-1}$ and a gain of 2.2
e$^-$ ADU$^{-1}$.

The area covered for the $U$ deep field consists of three pointings
partially overlapping (see Fig.\ref{Fig:field}). When this survey
started in 1999, a standard $U-$band Johnson filter was not available
at the ESO MPI 2.2m telescope, so our team borrowed the one available
at the Loiano observatory, and used this till a suitable $U-$band filter
was acquired by ESO. One pointing was therefore imaged using the Loiano
$U$ filter, which is a circular filter partially vignetting the field.
Pointings 2 and 3 were imaged with the ESO U/360 filter. 

\begin{figure}
\begin{center}\mbox{\includegraphics[%
  width=5cm,
  angle=270]{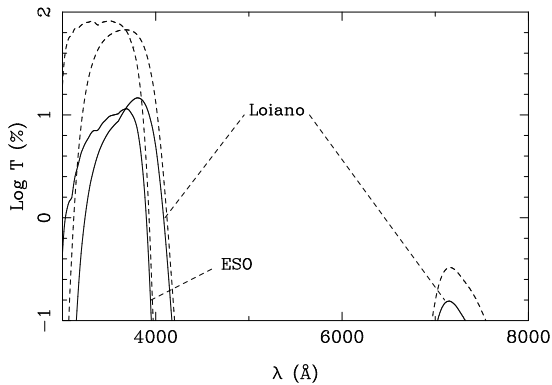}}\end{center}

\caption{Transmission curves of the Loiano and ESO U/360 filters; curves before
(dashed) and after (solid) the convolution with the CCD, telescope
and atmosphere transmission are displayed. Note the red leak for
the Loiano filter.}

\label{Fig:trasm}
\end{figure}

The filter transmission curves are shown in Fig.~\ref{Fig:trasm};
central wavelengths and FWHMs are given in Table~\ref{Tab:log}, together
with covered area and exposure times for each pointing. Note that,
compared to the ESO filter, the Loiano filter is shifted to the red
and also has a red leak at $\lambda\sim7000$\AA.

The efficiency during the observing campaign was hampered
by bad weather conditions and a problem at the primary mirror support
system which caused strong astigmatism on some images.

The exposure time for each individual frame was usually 2000s, 
and 1000s when astigmatism of the ESO MPI 2.2m telescope
was severe. The total
exposure time was 13.9 hrs for pointing P1, 20.3 hrs for pointing
P2, and 17.8 hrs for pointing P3; see Table~\ref{Tab:log} for additional
information on the observing log.

For each pointing, the sets of exposures were acquired in a dithered
elongated-rhombi pattern. This sequence of dithered exposures ensures
the removal of CCD gaps in the final coadded image, better flat fielding
correction, efficient removal of bad pixel and columns and a more accurate astrometric
solution in the final coadded image. The total area covered by the
three pointings is $\sim0.93 \deg^2$. The effective area
is actually smaller, $\sim0.71 \deg^2$, since: (a) a
fraction of P1 is overlapping with P2 and P3; (b) the border regions
of the images were masked when the noise was significantly higher
due to either vignetting (P1) or the dithering steps; (c) a small fraction
of the area in P2 and P3 is not covered in $BVRI$ and  therefore was not
taken into account when catalogs were built.

The atmospheric turbulence produced an average seeing of approximately
$\rm{FWHM}\sim1.3\arcsec$.

\begin{table*}
\begin{center}
\begin{tabular}{c cm{1.5cm} c cc cc cccc}
\hline 
Pointing& Date& Filter & \multicolumn{2}{c}{$\lambda_{c}$} & 
     \multicolumn{2}{c}{FWHM} & Area& Ditherings& Tot. exp & Seeing \\
        &     &        & \multicolumn{2}{c}{(\AA)}        
	& \multicolumn{2}{c}{(\AA)} 
	& ($\deg^2$)&     &  (hr)& (arcsec)\\
\hline
1 & Nov 99 & Loiano & 3620 & \emph{3750}& 527 & \emph{460}& 0.27& 25x2000s& 13.9& 1.4\\
2 & Oct 00, Nov 00 & ESO U/360 & 3404 & \emph{3540} & 732 & \emph{536}& 0.30& 
22x1000s+ 25x2000s& 20.3& 1.3\\
3 & Aug 01, Oct 01, Apr 02& ESO U/360& 3404 & \emph{3540}& 732 & \emph{536}&
0.29& 32x2000s& 17.8& 1.2\\
\hline
\end{tabular}\end{center}

\caption{Details concerning the pointings for the $U$ deep field. Filter central
wavelengths and FWHM are given before and after (in italics) the convolution
with the telescope, CCD and atmosphere transmissions. The area includes
only the region where sources have been included in the catalog: it
does not include border regions with vignetting or a low signal to noise
ratio.}

\label{Tab:log}
\end{table*}

\begin{figure}
\includegraphics[%
  clip,
  width=7cm,
  keepaspectratio]{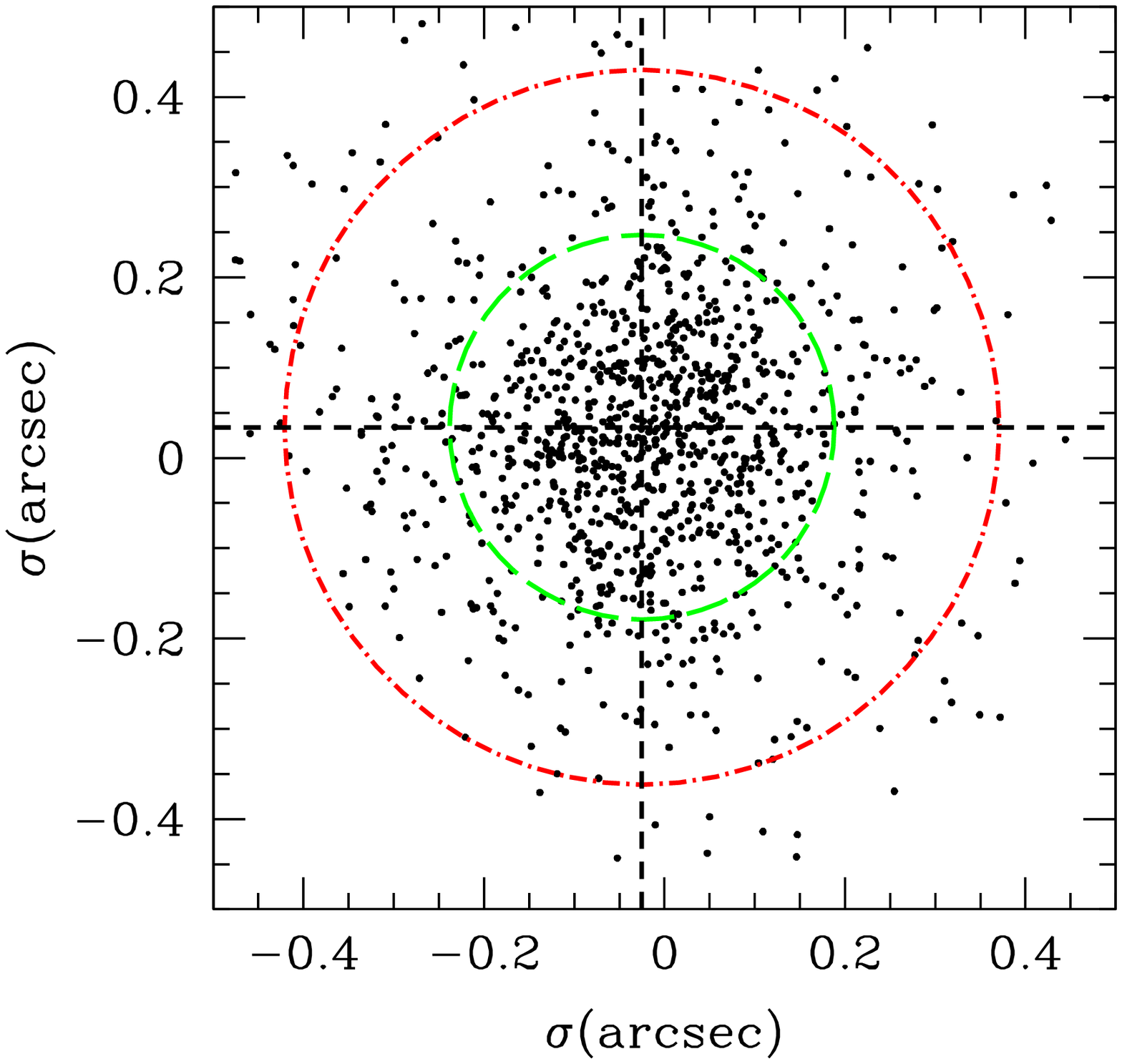}

\includegraphics[%
  bb=22bp 16bp 596bp 700bp,
  clip,
  width=6cm,
  keepaspectratio,
  angle=270]{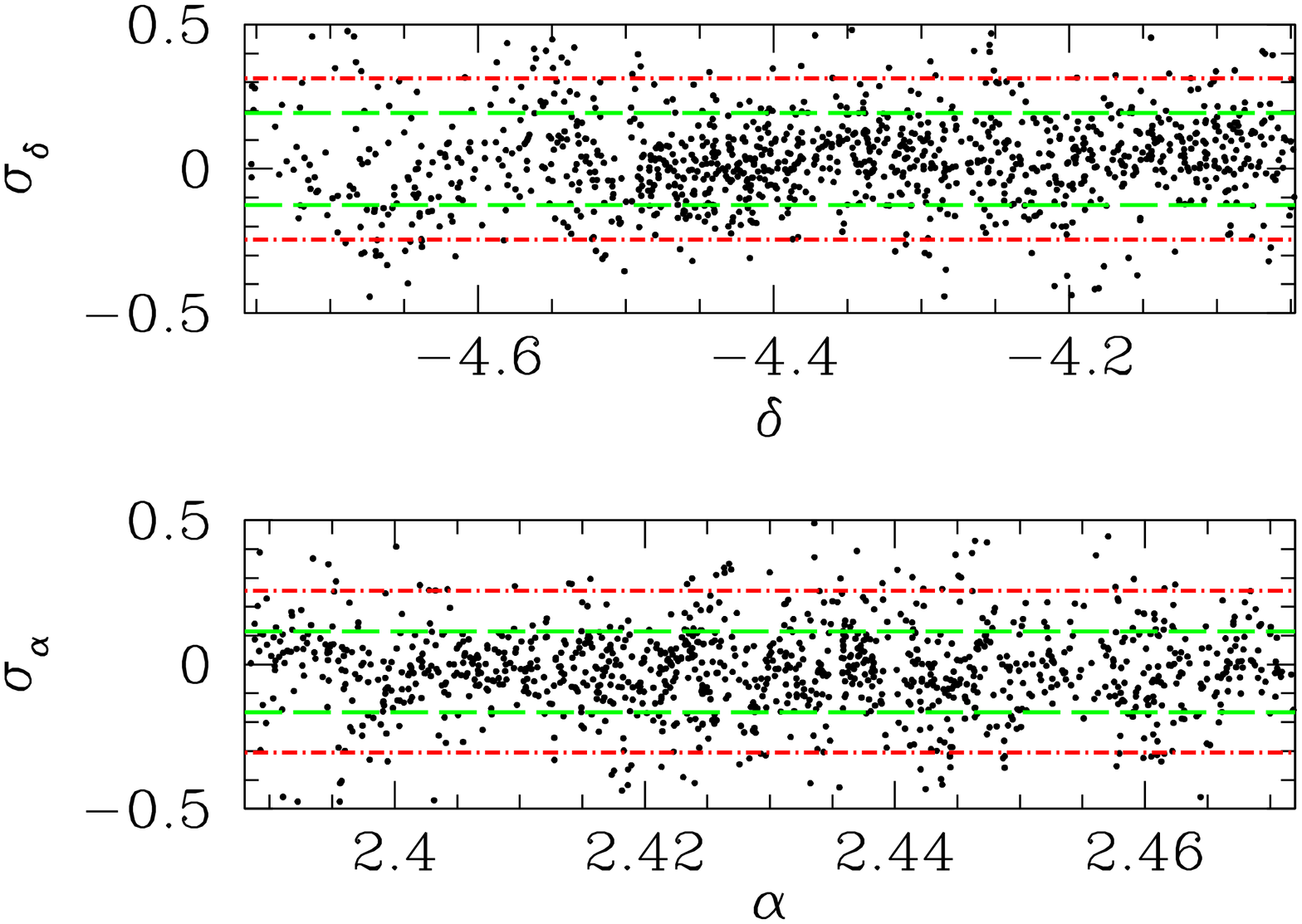}

\caption{Astrometric residuals for unsaturated, point-like sources over the
full field (P1, P2 and P3). Up: radial residuals; the inner and outer
circles enclose 68\% ($\sigma = 0.21\arcsec$) and 90\% ($\sigma = 0.48\arcsec$)
of the sources respectively; the cross shows the centroid, ($\sigma_{\alpha}$,
$\sigma_{\delta}$)$=$(-0.02\arcsec, 0.03\arcsec). Bottom: residuals are
displayed versus right ascension and declination; the lines display
the residuals per coordinate (68\%: $\sigma_{\alpha}=0.14\arcsec$, 
$\sigma_{\delta}=0.16\arcsec$; 
90\%: $\sigma_{\alpha}=\sigma_{\delta}=0.28\arcsec$). }

\label{fig:astro_rms}
\end{figure}

\subsection{Data reduction}

Pre-reductions were carried out using the {\sc MSCRED} package in {\sc IRAF}.
The CCD mosaic frames were bias and dark corrected, and flatfielded.
Flatfield images were constructed combining a whole series of twilight
sky images, taken during each observing night. Once the images were
flatfielded, we noticed the presence of residual structures in the
background sky, which should be treated in order to get a flat sky
background in the final coadded images. A {}``super-flatfield\char`\"{}
image was constructed from all the dataset for each pointing, using
a 3-$\sigma$ rejection algorithm, for the removal of sources
in the field. The procedures for astrometry, photometry and coaddition 
were approximately the same as those adopted in the analysis of $BVRI$ data.  
More details are given in Paper II; a short description follows where details
 peculiar to $U-$band data are emphasized.

\subsubsection{Astrometry}

Astrometry was performed using the {\sc astrometrix} tool, which is part of the
{\sc WIFIX}\footnote{The {\sc WIFIX} package was developed by M.R. in cooperation with the TERAPIX
team; it is freely available at http://www.na.astro.it/$\sim$radovich.
} package developed for the reduction of wide-field images.
{\sc Astrometrix} allows to compute an astrometric solution using both
an external astrometric catalog and the constraint that the position
of overlapping sources in different CCDs must be the same (global astrometry).

One of the main goals of this survey was to provide $U-$band fluxes
of the sources detected in the $BVRI$ images. As described in more detail
in Paper II, source detection  for the $BVRI$ images was done as follows.
A single $\chi^{2}$ image was first built from the $BVRI$ images, source
detection was then done using {\sc SExtractor} (Bertin \& Arnouts \cite{bertin})
in dual mode. This requires
that source positions in all the images match at a subpixel level.
Such level of accuracy was achieved as outlined in Paper II. The astrometric
solution was first computed for  the $I-$band image taking  the USNO A-2 (Monet et al. 1998)
as the astrometric reference catalog. A catalog of sources was 
then extracted from the resampled $I-$band image and used as reference catalog for
the other bands  ($UBVR$). 
During the astrometric procedure, the offset of $U-$band detected sources 
with respect to those matched in the $I-$band catalog was first computed for 
each CCD separately. 
This allowed to correct 
the displacement introduced by atmospheric refraction from the $I-$ to the 
$U-$ band. Finally, in the global astrometry step the astrometric solution
was constrained for each CCD by both the positions from 
the $I-$band catalog and those from overlapping sources in all the other CCDs.
Even if the three pointings were then resampled and coadded separately, in the global 
astrometry we used positions from all of them to increase the astrometric internal
accuracy by using the  position of the same source in the overlapping region of two different 
pointings (see Fig.~\ref{Fig:field}). 

To compute an accurate astrometry for
$U-$band data is somewhat more difficult than for $BVRI$ data because of the smaller number of
bright stars in the $U-$band. 
 This implies that the astrometric solution may heavily rely
on extended sources, for which the emission may peak at different positions
in different bands. The achieved RMS astrometric accuracy, measured using
$I-$band selected point-like sources only, is $< 0.2 \arcsec$ 
(see Fig. \ref{fig:astro_rms}). The {\em internal} RMS, computed from
overlapping sources in different exposures, is much smaller,
$\sim 0.07\arcsec$.

\begin{figure}
\includegraphics[%
  clip,
  width=8cm,
  keepaspectratio]{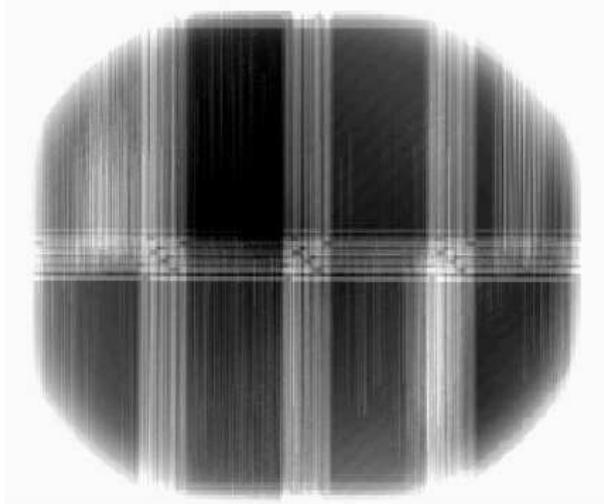}

\caption{Weight map produced after coaddition of the P1 pointing.\label{fig:weightP1}}
\end{figure}

\subsubsection{Photometry}

Photometric calibration was done taking several Landolt fields, so
that we had at least one Landolt star in each CCD. Landolt
fields were imaged also in $B$,$V$ to derive the filter color term, for
each $U-$band filter used.

The calibration equation becomes: 
\begin{equation}
U-u=\alpha(U-B)+\beta-\epsilon X,
\label{eq:photcal}
\end{equation}

Landolt fields were taken at different airmasses (\emph{X}) during
the night for all pointings with the exception of P2; 
\ub\ colors of the Landolt stars used in the calibration 
are in the range $0 \le \ub \le 1.5$.
This allowed  us to compute the extinction coefficient $\varepsilon$: a 
value close
to that expected for La Silla was obtained. For P2 this was not possible,
so we took the average La Silla extinction coefficient for that period.
Zero points, color terms and extinction coefficients computed for
each pointing are given in Table~\ref{Tab:cal}. Recent photometric
investigations (Manfroid, Selman \& Jones \cite{manfroid}) showed that the 
zero point changes across the WFI mosaic with a maximum difference of 
$\sim$ 0.08 mag as a consequence
of non-uniform illumination and scattered light.
As the Landolt fields are not wide enough to cover the whole WFI area and hence to compute
one zero point for each CCD independently, the zero points we obtained represent an average over 
the whole mosaic from the different observed Landolt fields. The photometric 
accuracy in our data is therefore limited to $\sim 0.05$ mag, corresponding 
 to about half the maximum difference in zero
points over the entire WFI mosaic.

 From  the knowledge of the system transmission
curves we then  computed the AB corrections, that are displayed
in the last column of Table~\ref{Tab:cal}.

\begin{table*}
\begin{center}\begin{tabular}{ccccccccc}
\hline 
Pointing&
 Date&
 Filter&
 $\alpha$&
 $\beta$&
$\varepsilon$&
 RMS &
\emph{X}&
 AB corr\\
\hline
1 &
 Nov 1999 &
 Loiano &
 0.21$\pm$0.02 &
 22.15 $\pm$ 0.06 &
0.43$\pm0.02$&
 0.06&
1.125&
 0.73\\
2 &
 Nov 2000 &
 ESO U/360 &
 0.08$\pm$0.02 &
 22.01 $\pm$ 0.02 &
0.49&
 0.06&
1.105&
 0.94\\
3 &
 Oct 2001 &
 ESO U/360&
 0.08$\pm$0.02&
 22.10 $\pm$ 0.02&
0.44$\pm$0.03&
 0.04&
1.139&
 0.94\\
\hline
\end{tabular}\end{center}

\caption{Zero points, color terms and airmasses for the
different pointings and filters. Date and airmass (\emph{X}) refer to the image taken as reference when
the relative photometry was computed. The last column gives the values to be
added to the Vega-system magnitudes to obtain AB magnitudes.}

\label{Tab:cal}
\end{table*}

To account for changes in airmass and atmospheric transparency, we
used the {\sc Photometrix} package in {\sc WIFIX}. This package first applies
the astrometric solution found for each CCD and then looks for overlapping
sources in different CCDs. It then computes \emph{for each mosaic}
(defined as the set of 8 CCDs taken for each exposure) an additional term to
the zero point ($z_{r,i}$) such that the average differences in fluxes
of overlapping sources are minimized. An exposure taken in photometric
conditions was chosen as reference, so that for it $z_{r,i}$=0: this
image was taken in the same night as the Landolt fields used to make
the photometry calibration. Since the P2 and P3 pointings are not
connected, and P1 was taken with a different filter, we needed to run
this step for each pointing separately.

\subsubsection{Coaddition}

After the astrometric solutions and flux scaling factors were computed,
image coaddition was done using the {\sc SWarp} tool developed by E. Bertin
\footnote{{\sc SWarp} is part of the TERAPIX software suite, available at 
http://terapix.iap.fr/soft}. 
{\sc SWarp} allows to subtract the background, resample the images according
to the astrometric solution, apply the flux scaling factor, and
finally combine them. As in the case of the $BVRI$ images, resampling
was done using a "Lanczos-3" interpolation kernel, which corresponds to a sinc function
multiplied by a windowing function.
The coaddition was
done computing the median of the images to optimize the rejection
of spurious sources as cosmic rays or satellite tracks.
As a consequence of the use of two different filters for the P1 and 
P2, P3 pointings, it was not possible to produce a single coadded image for the whole field. 
We therefore produced one image for each pointing; the size, pixel scale 
(0.205$\arcsec$ pix$^{-1}$) and orientation of each
image was  the same as that of $BVRI$ data so that catalogs can be extracted with the
same $\chi^2$ technique (see Paper II). This
allows to extract catalogs using {\sc SExtractor} in dual mode, with the
$\chi^{2}$ image as reference. In {\sc SWarp}, each image may be associated
to a weight-map to properly weight pixels during the coaddition step.
This was particularly useful in the case of the Loiano filter, due
to vignetting in the outer regions. Weight maps were first created
using normalized flat fields; pixels flagged in the Bad Pixel Maps
provided for WFI by ESO\footnote{http://www.eso.org/science/eis/eis\_soft/soft\_index.html} 
 were then set to 0 in the weight map. 
For the pointing P1, we also set to 0 those pixels 
with a value $< 0.6$ in the normalized flat field, thus allowing
us to remove in the coaddition the regions affected by vignetting. 
Fig.\ref{fig:weightP1} shows the coadded weight map for P1.

\begin{figure}
\begin{center}\includegraphics[%
  width=8cm,
  keepaspectratio]{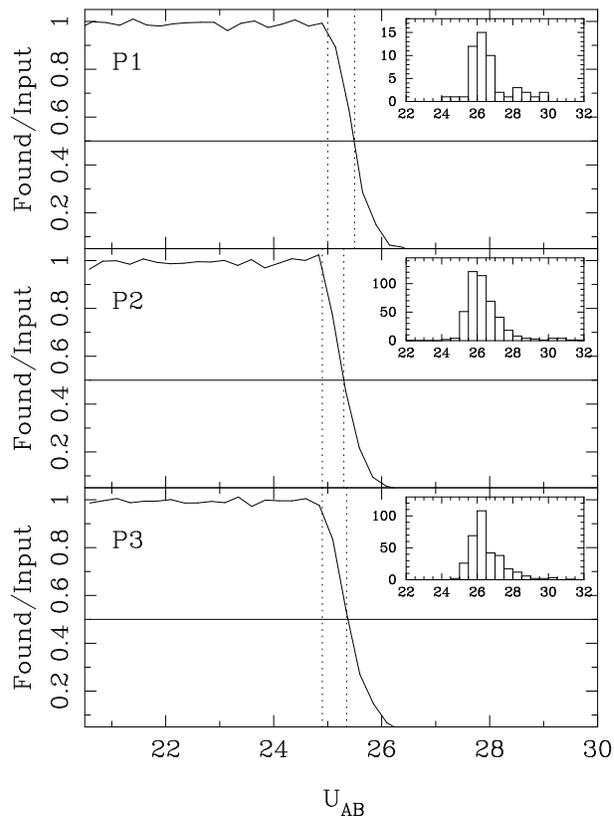}\end{center}

\caption{Ratio of detected to input sources as a function of input magnitude: 
the input catalog consists of simulated point-like sources with a flat 
magnitude distribution. The dotted lines show the adopted 50\% and 90\%
completeness magnitudes for each pointing. 
The histograms in the insets show the magnitude 
distribution  of the spurious sources detected
in the images with background only and no simulated
sources.}

\label{Fig:magcomp}
\end{figure}

\section{Preparation of the catalogs}

\subsection{Photometric properties\label{photU}}

A first estimate of the detection limiting magnitude for point-like sources was done 
on the background RMS map: this map was extracted for each pointing using {\sc SExtractor} 
and the median background RMS ($\sigma$) was then computed. The limiting 
magnitude at 3$\sigma$ and 5$\sigma$ levels is 
mag$(n\sigma) = \beta -2.5 \log(n \sigma \sqrt{A})$ where $n = 3, 5$,
$A$ is the area of an aperture whose radius is the average
FWHM of point-like sources, 
$\beta$ is the zero point  (Eq.~\ref{eq:photcal}). 
We obtained $U_{AB} \sim$
25.8 at 5$\sigma$ and $U_{AB} \sim$ 26.4 at 3$\sigma$. 

Because of their definition, these values for the limiting magnitudes
refer to the unrealistic case of objects with a flat radial surface
brightness profile.
A more accurate determination of the photometric limits of the survey
was then done as follows.
Catalogs were first  extracted using {\sc SExtractor}, where 
source detection was optimized using simulations of point-like sources. 
A background image was computed as outlined in Arnaboldi
et al. (2002); a population of 2000 point-like objects with the same
PSF of bright isolated stars in the field was added on this image,
following a flat magnitude distribution (20.5 $\leq$ mag $\leq$ 27)
and a random spatial distribution: the tasks in the {\sc ARTDATA} package in 
{\sc IRAF} were used. 
{\sc SExtractor} was then run on that
image and it was measured how many objects were detected and the faintest
magnitude reached for a given detection threshold. 
The number of connected pixels required for source detection was 
set to 9 pixels: this gives a  minimum signal to noise ratio of 
$\sim$ 3$\sigma$ for a detection threshold $\sim$ 1.
The same process
was repeated 20 times, giving a total of 40\,000 sources and around
1500 sources per 0.25 magnitude bin. At the same time, we monitored
the number of spurious sources, by matching the output catalogue from
{\sc SExtractor} with the input catalogue. 
The optimal threshold was the one  that minimized the number of spurious
detections with $U_{AB} < 26$, without loosing  input sources.
Following these tests, we set the detection threshold  to 
$\sim 0.7 \sigma$ for the P1 pointing and $\sim 0.9\sigma$ for P2 and P3, 
 over at least nine connected pixels.

Fig.~\ref{Fig:magcomp} shows the results of the simulations 
(detected/input sources). 
In Table~\ref{Tab:mags}
we list the magnitude of completeness ($>$ 90\% of the modeled objects
in the input catalog are retrieved) and the limiting magnitude (50\%
of the modeled objects retrieved). 
The depth of P1 is comparable to that
of P2 and P3 even if the exposure time is lower, since the Loiano filter
is redder than the ESO filter and is thus somewhat more efficient, after
convolution with the atmosphere. 
In order to check how many spurious detections should be expected
at different magnitudes, we then run {\sc SExtractor} on the background 
image before simulated sources were added. The result is displayed in the
histograms in  Fig.~\ref{Fig:magcomp},  which show that
there are  no spurious detections for magnitudes brighter than 
the 90\% completeness magnitude; the peak occurs 
for magnitudes fainter than the detection limit ($U_{AB} \ge 26$) 
\footnote{Spurious detections with magnitudes in the range 
$28 < U_{AB} < 30$ are mainly due to correlated noise.}. 
Residual spurious detections for magnitudes close to the 50\% limiting
magnitude are later removed by the cross correlation with the
catalog derived from the  $\chi^2$ image.

\begin{table}
\begin{center}\begin{tabular}{cccc cc}
\hline 
Pointing &
&
$m_{\rm comp}$ &
$m_{\rm lim}$ &
$\mu_{\rm comp}$ &
$\mu_{\rm lim}$  \\
 & &  (90\%) & (50\%) &  (90\%) & (50\%)\\
\hline
1&
&
 25.0 &
 25.5 &
 25.7 &
 26.2 
\tabularnewline
2&
&
 24.9 &
 25.3 &
 25.8 &
 26.3 
\tabularnewline
3&
&
 24.9 &
 25.3 &
 25.7 &
 26.2  
\tabularnewline
\hline
\end{tabular}\end{center}

\caption{Completeness and limiting magnitudes  and surface brightnesses
(mag arcsec$^{-2}$) derived from simulated  point-like and extended sources 
respectively. All values are in the AB system.}

\label{Tab:mags}
\end{table}

\begin{figure}
\begin{center}\includegraphics[%
  scale=0.5]{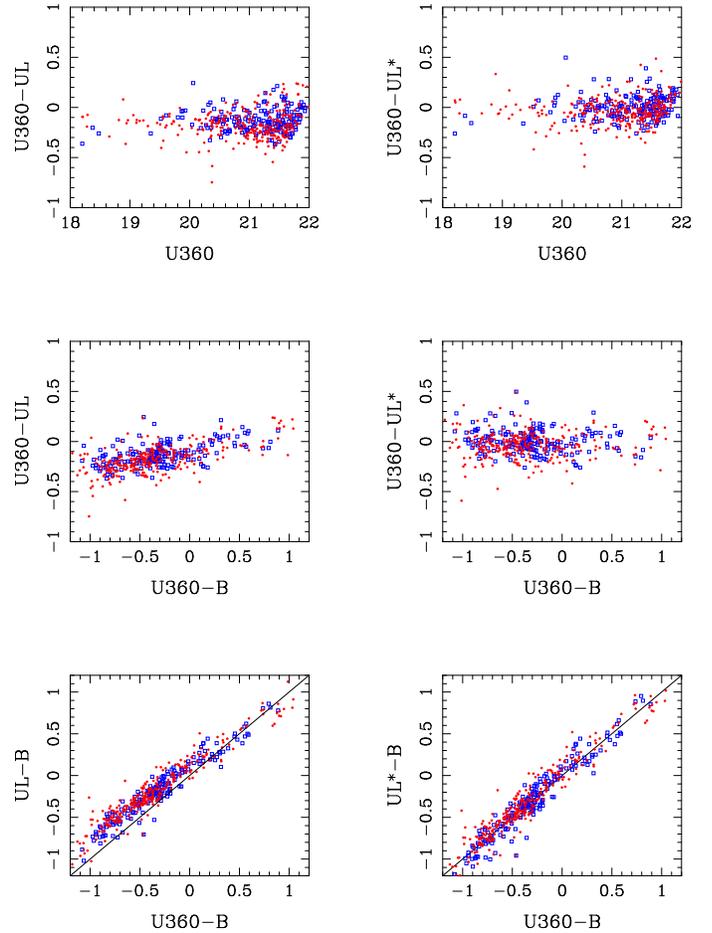}\end{center}

\caption{Difference in aperture magnitudes (5\arcsec diameter) of 
point-like and extended bright ($18 < U < 22$) sources taken with both the 
ESO (U360) and Loiano (UL) filters. 
$B-$band fluxes were taken from the VIRMOS  catalog.
The upper and middle panels show the differences in magnitudes before (left)
and after (right) the color correction is applied (squares: P2; dots:
P3). In the bottom panel $\ub$ colors from the Loiano and ESO filters
are compared. }

\label{Fig:magcc}
\end{figure}

Similar simulations were done using extended sources, to compute the 
completeness and limiting surface brightness. 
Galaxies with  De Vaucouleurs and exponential surface brightness laws were 
used as input sources to simulate elliptical and spiral galaxies respectively;
total magnitudes of these sources were in the
range $20 < U_{AB} < 26$. The same procedure as above was then followed, but
this time we measured the peak surface brightness (defined as the surface 
brightness of the brightest pixel) of the simulated and detected sources. 
The so obtained values are also given in Table.~\ref{Tab:mags}.

\subsection{Catalog extraction}

In order to easily combine the catalogs derived from the $U$ and $BVRI$
images, it was decided to use the $BVRI$ $\chi^{2}$ image catalogue
as reference for the extraction of sources in the $U-$band.

The $BVRI$ $\chi^{2}$ image catalogue is significantly deeper than the 
catalogue which can be derived using only the $U-$band image; the relative 
depths of
the two catalogues can be judged by comparing the surface densities of
objects, $\sim  10^5$ per square degree in the $\chi^{2}$  image catalogue 
(Paper II), and $\sim  10^4$ per square degree in the $U$ band catalogue. 
On the other hand, we would measure positive (spurious)
flux in the background-subtracted $U$ frame at the position of sources detected
in the $\chi^{2}$  image even if no significant $U$ source is detected (by
definition a positive flux would be measured in about 50\% random positions).
To remove these spurious detections we adopted the following strategy. 
For each pointing (P1, P2 and P3), we extracted two catalogs:

\begin{itemize}
\item One catalog was extracted using {\sc SExtractor} in the ``dual image
mode'', taking as input catalogue the one obtained from the $\chi^{2}$
image of the $BVRI$ dataset (see Paper II). 
\item One catalog was extracted in single-image mode, using a filtering
Gaussian kernel and the detection threshold set   as discussed above. 
\end{itemize}
For both catalogs,  Kron ({\sc magauto}) magnitudes and aperture magnitudes
at diameters of 15 pixels ($\sim 3\arcsec$) and 25 pixels ($\sim 5\arcsec$) 
were computed. As it concerns the required number of 
contiguous pixels, we used the same value that was adopted for the $BVRI$ 
images (see Paper II).

A cross-correlation was then done between coordinates of sources in
the two catalogs; 
a maximum distance of $2\arcsec$ was adopted for matching.
 In addition, aperture magnitudes ($5\arcsec$ diameter) were also 
used  as a further constraint in the cross-correlation: matching sources 
were rejected when the difference of $U-$band magnitudes obtained from  
the $\chi^2$ and single image catalogues  was $> 0.5$ mag.

 After the cross correlation we computed the average difference between the
 magnitudes measured in the single and $\chi^2$ images: for all the 
three pointings we obtain $\langle\Delta U\rangle \le  0.02$ mag with RMS 
values which change from $< 0.05$ mag
for bright sources ($U_{AB} < 22$) to $\sim  0.1$ mag for fainter magnitudes.
Considering (1) the different depths of the $U-$band and $\chi^2$ images
and (2) the fact that for extended sources the peak of the emission may be 
different in the $U-$band compared to the $BVRI$ bands, we 
decided to keep in the final catalog the magnitudes measured on the single 
image rather than those computed using the parameters from the $\chi^2$ image.

As a consequence of the dithering strategy, regions close to the border
of the image are covered by a small number of exposures: spurious sources
are therefore produced either by a bad rejection of cosmic rays or
by the high noise. We therefore removed from the catalogs all the
sources located within a given distance from the border. $U-$band magnitudes
were assigned to sources in the $\chi^{2}$ catalog only when they
were detected in the single-image catalog with a magnitude brighter than 
the $3\sigma$ limiting magnitude, $U_{AB} = 26.4$. 
For magnitudes brighter than the completeness limit, unmatched sources
are mainly false detections (e.g. halos around bright stars) or sources
not found in the $\chi^{2}$catalog (e.g. because of cosmetic defects).
For fainter magnitudes, spurious detections due to correlated noise are more
important in the single-image catalog (see Fig. \ref{Fig:magcomp})
but are removed by the cross correlation with the $\chi^{2}$ catalog.
The requirement on the magnitude helps to reject these spurious 
detections, removing $\le$ 6\% of sources for $U < 25$ mag, $\sim$ 10\% for 
fainter magnitudes.

The separation of extended vs. point-like sources was done on the $I-$band
catalog, and is described in Paper II: such classification was done
in the range $18 < I_{AB} < 21$, that is between the $I-$band
saturation limit and the magnitude beyond which the separation between resolved
and unresolved sources is less reliable. We verified on the half-light
radius versus $U-$band magnitude plots that sources classified as point-like
fall on the expected locus. The same plots were also used
to estimate the average seeing in each pointing that is given in 
Table~\ref{Tab:log}.
In addition, bright nearby galaxies
which are saturated in the $I-$band but not in the $U-$band were
flagged as extended.

Finally, we computed for each filter the correction to bring the magnitudes
to the AB system (see Table~\ref{Tab:cal}) and the Galactic extinction
correction. The F02 field is characterized by a low interstellar extinction 
(see Paper I); according to the Schlegel et al. 
(\cite{schlegel}) maps\footnote{The dust map was downloaded from: \\ 
http://astron.berkeley.edu/davis/dust/data/data.html.} 
$0.023 < E\bv < 0.040$,  
with an average value $E\bv \sim 0.027 \pm 0.002$ mag.
This translates to an average correction for extinction
$A_U \sim 0.14$ mag. 

\begin{figure}
\includegraphics[%
  width=6cm,
  keepaspectratio,
  angle=270]{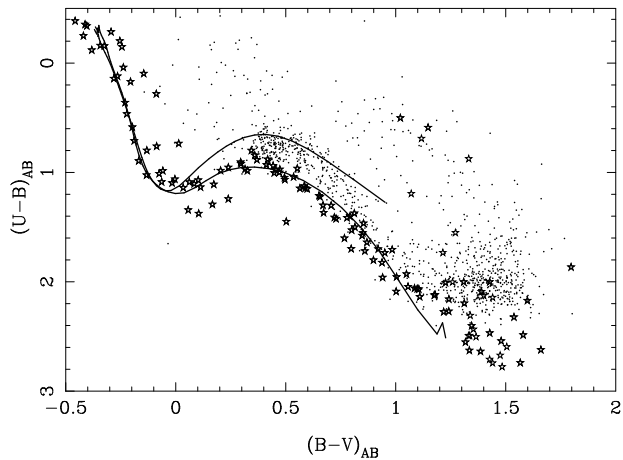}

\caption{Comparison of colors for bright ( $U_{AB} < 23.5$)  point-like 
sources (dots) with:  the library from Pickles (1998) 
(\textit{stars}); 
the Kurucz model atmospheres (Kurucz \cite{kurucz}) with $\log g=4.0$,
{[}M/H{]}=0 (lower curve) and {[}M/H{]}=-5.0 (upper curve). Magnitudes
in P1 were trasformed to the ESO filter photometric system.}

\label{Fig:starcol}
\end{figure}

\begin{figure}
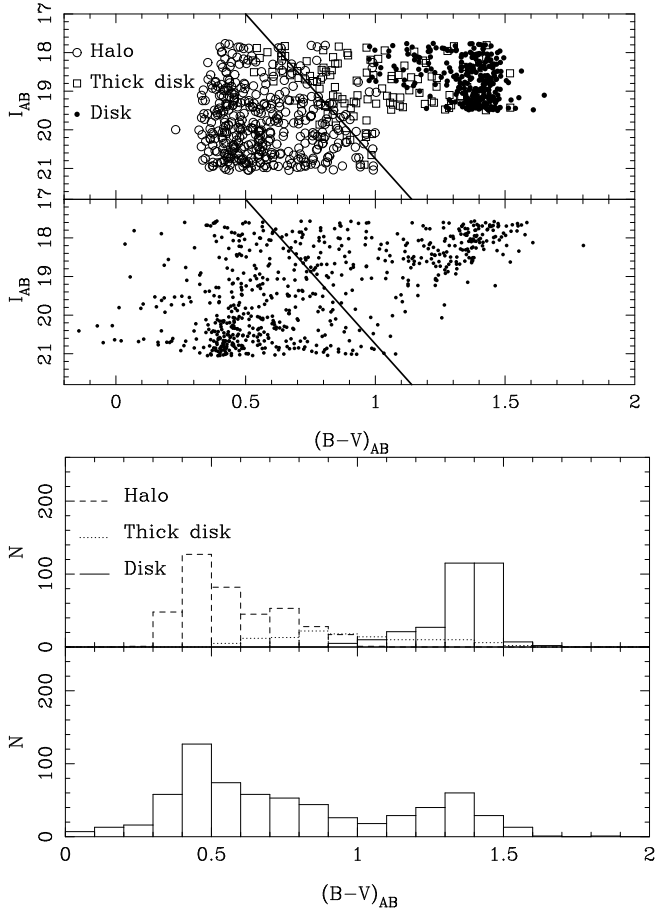

\includegraphics[%
  width=6cm,
  keepaspectratio,
  angle=270]{figure8a.ps}
\includegraphics[%
  width=6cm,
  keepaspectratio,
  angle=270]{figure8b.ps}

\caption{Comparison of simulated and  observed $\bv$ colors vs. $I$
magnitudes. {\em Top:} color-magnitude diagram. The upper panel
shows the synthetic colors expected in the F02 area.
The lower panel shows the observed colors for $U-$band selected 
($U_{AB} < 23.5$) point-like sources  from the three pointings; the cuts
at the faint and bright end are due to the fact that the selection
of point-like vs. extended sources was done in the range $18 < I_{AB}
< 21$  only. The solid line indicates the separation 
between halo and disk/thick disk stars. 
{\em Bottom: } The same as above, but plotted as a histogram. 
\label{Fig:bvcheck}}
\end{figure}

\begin{figure}
\includegraphics[%
  width=7cm,
  angle=270]{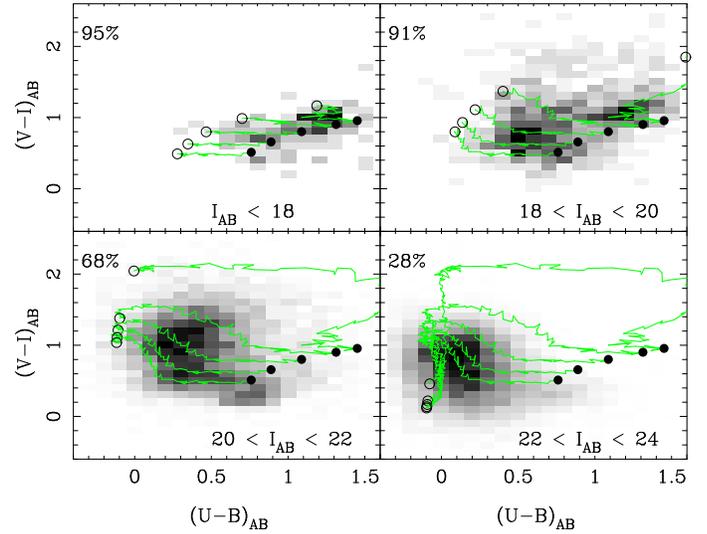}

\caption{$\ub_{AB}$ vs. $\vi_{AB}$ colors for extended sources
with $U_{AB}< 25$. Colors are displayed as a grey-scale map
where the level is proportional to the number of sources in a given
color bin. Model tracks for early to late type galaxies  are also displayed
(starting from right to the left);
the positions of lower and upper redshift are indicated by the filled
and open circles respectively. Magnitudes in P1 were transformed to
the ESO filter photometric system. The four panels show sources in
different bins of $I_{AB}$, as follows. (a) $I_{AB} <18$, 
model tracks: $0.0 < z < 0.2$. (b) $18 < I_{AB} < 20$, model
tracks: $0.0 < z < 0.4$. (c) $20 < I_{AB}< 22$, model tracks:
$0.0 < z < 0.8$. (d) $22 < I_{AB}< 24$, model tracks: $0.0 < z
< 1.5$. The value in  the top left corner is the fraction of galaxies 
detected in the $U-$band with respect to those detected in the $I-$band, 
in each bin. 
\label{Fig:galcol}}
\end{figure}

\subsection{Color correction of the Loiano filter to the ESO filter}
\label{subsec:intcorr}

After catalogs were extracted, for each pointing we assessed the problem of bringing
the $U-$band photometry to a single  photometric system. We decided to convert
the magnitudes measured in P1 with the Loiano filter to the ESO photometric system,
 since the latter is closer to the Johnson system. The advantage of doing so,
rather than using  Eq.\ref{eq:photcal} to transform both magnitudes to the Johnson system
and then computing the internal offset, is given by the fact that we can later use the 
ESO system transmission curve e.g. for comparison with model predictions.

As shown in Fig.\ref{Fig:field},
the P2 and P3 fields taken with the ESO filter partially overlap with
P1. For the overlapping sources, we therefore
have photometry for both the $U-$band filters; in addition, $B-$band
photometry is available from the CFHT data.
We first extracted from the catalog bright ($20 < U < 22$) point-like sources; 
35 and 46 sources were found for P2 vs. P1 and P3 vs. P1 respectively.
From these we solved Eq.\ref{eq:photcal} with no airmass ($\epsilon=0$),
taking the ESO filter as reference (magnitudes not including the AB
correction were used).
Similar results were obtained for both P2 and P3: hence  we merged the
two sub-catalogs and finally obtained $\alpha=0.178$, $\beta=-0.08$ in 
Eq.\ref{eq:photcal}. The
color term, as expected, is intermediate between the values found for the two filters
from the calibration with standard
stars (see Table \ref{Tab:cal}). In Fig.\ref{Fig:magcc} we show the
differences in magnitudes of both point-like and extended sources in
the overlapping areas of the two filters before and after color correction.
The average offset in magnitude after the correction is -0.01 mag (P1 vs. P2) and
0.04 mag (P1 vs. P3), with an RMS $\sim 0.1$. 
The same plot also provides an indirect check that photometry in P2
and P3 is consistent, as each of them is consistent with the P1 photometry
in the overlapping areas. 

The consistency with magnitudes obtained in the Johnson system by 
applying the color terms listed in Table \ref{Tab:cal} was checked as follows. 
Both the ESO and Loiano magnitudes were first transformed to the 
Johnson system. The same procedure as above was 
then followed to compute the offset between the Loiano and ESO photometry: 
we found  $\alpha=0$ as expected and $\beta=-0.1$ mag. Loiano magnitudes
were then transformed to the ESO system as discussed above, and then to the
Johnson system. The agreement is within 
$\sim 0.08$ mag in the range $-2 < \ub <2$.

\section{Data quality assessment}

In order to check the data quality of the $U-$band photometry, we performed
a series of tests where we (i) compared colors of point-like and extended
sources to those expected for stars and field galaxies; (ii) checked
the number counts of galaxies and compared them with existing data in 
literature.  In this analysis, magnitudes in the P1 pointing  were first 
transformed to the ESO filter photometric system as described 
in Sec.~\ref{subsec:intcorr}; for those sources which were  
observed with both the Loiano and the ESO filter, the latter magnitude
data were taken. In the final catalog, magnitudes are corrected for 
Galactic extinction.

\subsection{Stellar colors}

We first compared the colors of bright ($U_{AB} < 23.5$)  point-like sources
with those obtained from the convolution of stellar templates (Pickles
\cite{pickles}) with the system transmission curves (filter + telescope
+ CCD + atmosphere). No correction for Galactic extinction is applied
in this case.
The result is displayed in Fig.~\ref{Fig:starcol}:
it can be seen that the agreement is good only for $\bv \geq$ 0.7. This
is expected since for bluer colors the low-metallicity stellar population
from the halo dominates over the disk population with solar metallicity.
We therefore computed the colors produced using the Kurucz model atmospheres
(Kurucz \cite{kurucz}) with $\log g=4.0$ and {[}M/H{]}=0.0, -5.0
to reproduce solar and sub-solar metallicities (see also Lenz et al.
\cite{Lenz}). As displayed again in Fig.~\ref{Fig:starcol}, the
apparent excess in $\ub$ is in agreement with the colors produced by the 
low-metallicity model. 
The absence of observed stars at redder colors ($\ub > 2.5$) is due to 
the cut in the $U-$band magnitudes, as it is explained later in more detail. 
Most of the sources outside the
stellar locus with an ultraviolet excess ($\ub < 0.6$) are likely to be 
quasars at $z <  2.2$. 

In order to verify whether the color distribution we obtained is consistent
with that expected in our Galaxy, we show in Fig.\ref{Fig:bvcheck}
(top) the synthetic color-magnitude (CM) diagram ($\bv$ vs. $I$) for a region 
of sky centered on P1 and a size of 0.9 $\deg^2$ (Degl'Innocenti \& Cignoni,
Private Communication). The code and recipes described in 
Castellani et al. (\cite{castellani}) were used to this aim: the code provides 
$BVRI$ magnitudes for a Galactic stellar population model, including 
halo, disk and thick disk components.
 
We selected from the simulations those sources with  
$18 < I_{AB} < 21$ as this is the range where the point-like classification
was done in our data. 
 We expect to find only halo stars (i.e. metal
poor objects) around $\bv \sim 0.5$, a mix of halo and thick disk stars
in the interval $0.5< \bv <1.4$, and mainly disk stars (metal rich
objects) around $\bv \sim 1.4$. 
Fig.\ref{Fig:starcol} shows that $\ub > 2$ when $\bv > 1$: 
therefore, when we compare simulated and observed color distributions we need 
to take into account the selection  introduced by the cut in the 
$U-$band magnitude,  $U_{AB} < 23.5$.
As the models do not provide
$U-$band magnitudes, we proceed as follows. The limits $\ub > 2$ and
$U_{AB} < 23.5$ imply that $B_{AB} < 21.5$. From $\bv >  1$ we obtain 
$V_{AB} < 20.5$. Finally, Fig.~15 in Paper II shows that 
$\vi > 1$ when $\bv > 1$: we therefore obtain $I_{AB} < 19.5$. 

The same diagram is plotted in 
Fig.\ref{Fig:bvcheck} for the point-like sources in the VIRMOS deep
field which were detected  in the $U-$band.  An exact match 
with the number of disk stars can not be expected as the selection 
due to the cut in the $U-$band was taken into account in an approximate way.
The position of the two clumps 
centered at $\bv \sim 0.5$ and $\bv \sim 1.4$ is clearly  visible 
(Fig~\ref{Fig:bvcheck}, bottom). This confirms that 
the $\ub$ excess seen in Fig.~\ref{Fig:starcol} is due to the halo stars in the
F02 field.

\begin{figure}
\includegraphics[%
  width=6cm,
  keepaspectratio,
  angle=270]{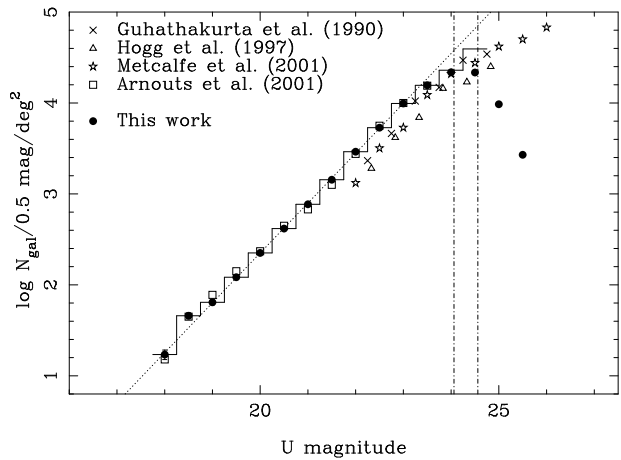}

\caption{Galaxy magnitude-number counts from the VIRMOS $U-$band deep field
compared with existing literature data. $U-$band magnitudes are in Vega system.
The filled circles show the VIRMOS data 
(statistical error bars are displayed but are too small to be seen);
magnitudes from P1 were transformed to the ESO filter photometric
system. Literature data are taken from the N. Metcalfe's
collection. The dotted line shows the best
fit computed in the range $18.5 < U < 22.5$. The dot-dashed lines show the
90\% and 50\% completeness limits. The histogram displays the number counts
corrected for incompleteness (see text for details).}

\label{fig:counts}
\end{figure}

\subsection{Field galaxy colors}

$\ub_{AB}$ and $\vi_{AB}$ colors were 
compared with those obtained from model galactic spectra convolved
with the system transmission curves. This was done using the {\sc make\_catalog}
tool which is part of the {\sc HYPERZ} package (Bolzonella et al. \cite{bolzonella}):
we used the built-in set of templates for early to late- type galaxies
computed from the GISSEL98 spectral evolution library of Bruzual \&
Charlot (\cite{bruzual}). As in Paper II, we divided our catalog
of $U-$band detected extended sources in  different bins of $I_{\rm AB}$ 
magnitude. Late-type galaxies are expected to increasingly
dominate in the color distribution as we go to fainter magnitudes:
early-type galaxies with faint $I$ magnitude are too faint to be included 
in the $U-$band catalog.  

\noindent (i) $I_{AB}<18$. The brightest galaxies are clearly
dominated by a low redshift population, $z < 0.2$, of early to late-type
galaxies (Fig. \ref{Fig:galcol}a).

\noindent (ii) $18 < I_{AB}< 20$. In Paper II it is shown
that $\bv$ vs. $\vi$ colors for these galaxies are in agreement with
those expected for a population of galaxies with $z$ = 0.0 - 0.5. The
median redshift expected for this magnitude range is $z \sim0.2$, 
according to spectroscopic surveys like the CFRS (Crampton et al. 
\cite{crampton}).
Consistent results are found looking at the $\ub$ vs. $\vi$ colors
(Fig. \ref{Fig:galcol}b), where early to late type galaxies are all
seen to contribute to the observed colors. 

\noindent (iii) $20 < I_{AB}< 22$. The colors
are well reproduced by a population  dominated by late type
galaxies with $z \sim$ 0.0 - 0.8, with a peak at $z \sim 0.35$ 
(Fig. \ref{Fig:galcol}c).

\noindent (iv) $22 < I_{AB}< 24$. The limiting magnitude
$U_{AB}$ $\sim$25 implies that in this range we mainly
select galaxies with bluer colors ($\ub_{AB} < 0.2$): 
most of these are likely to be late type galaxies at $z < 1$,
with some contribution from higher redshift galaxies 
(Fig. \ref{Fig:galcol}d). Note that in this magnitude range
less than 30\% of the galaxies are detected in the
$U-$band.

\subsection{Number counts}

The comparison of number counts of galaxies with literature data provides
a good check for both  the efficiency of the star-galaxy separation
(done in the $I-$band in our case) and the quality of our photometry.
For $U-$band data, this is complicated by the few published number counts
and by the heterogeneity of the data (e.g. different photometric systems,
no correction for galactic reddening). A collection of $U-$band number
counts is provided in the Vega-system by N. Metcalfe's  
(http://star-www.dur.ac.uk/$\sim$nm/, see also
Metcalfe et al. \cite{metcalfe}); we selected only those measurements
from CCD observations. They are displayed in Fig.\ref{fig:counts}. 
The offset among the different measurements is due to both the different
$U-$band filters and the absence of reddening correction for some of the
literature number counts. For example, in the case of  Metcalfe's WHDF data 
$E\bv_{WHDF} \sim 0.02$, so that after dereddening the WHDF number counts 
should be shifted by  $\sim$ -0.1 mag (Heidt et al. \cite{heidt}).

The $U-$band VIRMOS number counts  were normalized to the area covered by the 
unmasked regions, taking into account the overlapping regions  
($\sim 0.7\deg^2$). Number counts not corrected for incompleteness are 
displayed as filled circles in Fig.\ref{fig:counts}. 
Our data are in very  good agreement with those of Arnouts et al.
(\cite{arnouts}), which were also taken with ESO/WFI. A least-squares
fit in the range $18.5 < U <22.5$ gives a slope 
d(log N)/dm = 0.54$\pm$0.06.
The histogram in Fig.\ref{fig:counts} shows the number counts corrected
for the ratios of input to detected sources displayed in Fig.~\ref{Fig:magcomp}
(Sec.~\ref{photU}): after this correction,  number counts are in agreement 
with the extrapolation from  brighter magnitudes up to $U \sim 24.5$ 
($U_{AB} \sim 25.5$).

\section{Summary}

$U-$band data obtained in the framework of the VIRMOS preparatory imaging 
survey for the F02 deep field were presented in this paper, as a complement
to the $BVRI$ data presented in Paper II. Observations, data reduction
and catalog extraction issues were first discussed. Various quality 
assessment tests were then performed. A good agreement is found between 
the observed stellar colors and those computed from  the  
Kurucz model atmospheres;
a component with sub-solar metallicity is required to fit the halo stellar
population which dominates over the disk population for $\ub < 1.5$.
This is confirmed by the comparison with the color distribution of stars
in our Galaxy based on the Castellani et al. (\cite{castellani}) model. 
The colors of extended sources were compared with those obtained from
template spectra of galaxies. These tests also allowed to  check the 
photometric consistency
of $U-$band and $BVRI$ data. Number counts for extended sources were 
finally computed
and compared with other $U-$band data in literature. 

\begin{acknowledgements}
We wish to thank Scilla Degl'Innocenti for having kindly provided to us the
simulated data for the Galactic stellar population. We also thank
Fernando Selman and the ESO La Silla 2.2m team who sent us the transmission 
curve of the Loiano filter. 
Data analysis and plots were done using the Perl Data Language
(PDL), which is freely available from \char`\"{}http://pdl.perl.org\char`\"{};
PDL is a powerful vectorized data manipulation language derived from
perl.  This paper made
use of the NASA/IPAC Extragalactic Database (NED) which is operated
by the Jet Propulsion Laboratory, California Institute of Technology,
under contract with the National Aeronautics and Space Administration.
M.R., Y.M. and E.B. were partly funded by the European RTD contract
HPRI-CT-2001-50029 ''AstroWise''. Part of this work was also 
supported by the Italian Ministry for University and Research (MURST) under
the grant COFIN-2000-02-34. 

\end{acknowledgements}

\end{document}